# THz spectroscopy on amino acids

Sebastian Emmert, Peter Lunkenheimer, and Alois Loidl[a]

*Experimental Physics V, Institute of Physics, University of Augsburg, 86135 Augsburg,* Germany

**Abstract**

We present a detailed study on the temperature-dependent THz spectra of the polycrystalline amino acids L-serine and L-cysteine for wave numbers from 20 to 120 $cm^{-1}$ and temperatures from 4 to 300 K. Even though the structure of these two amino acids is very similar, with a sulfur atom in the side chain of cysteine instead of an oxygen atom in serine, the excitation spectra are drastically different. Obviously, the vibrational dynamics strongly depend on the ability of cysteine to form sulfur-hydrogen bonds. In addition, cysteine undergoes an order-disorder type phase transition close to 80 K, with accompanying anomalies in our THz results. On increasing temperatures, well-defined vibrational excitations, exhibit significant shifts in eigenfrequencies with concomitant line-broadening yielding partly overlapping modes. Interestingly, several modes completely lose all their dipolar strength and are unobservable at ambient conditions. Comparing the recent results with published work utilizing THz, Raman, and neutron-scattering techniques, as well as with ab-initio simulations, we aim at a consistent analysis of the results ascribing certain eigenfrequencies to distinct collective lattice modes. We document that THz spectra can be used to fine-tune parameters of model calculations and as a fingerprint property of certain amino acids. In addition, we analyzed the temperature-dependent heat capacity of both compounds and detected strong excess heat capacities at low temperatures compared to the canonical Debye behavior of crystalline solids, indicating soft excitations and a strongly enhanced phonon-density of states at low frequencies.

[a]Author to whom correspondence should be addressed: alois.loidl@physik.uni-augsburg.de

## I. Introduction

THz radiation, covering a frequency regime from 0.1 – 10 THz (1 THz ~ 33 cm$^{-1}$ ~ 4.1 meV ~ 48 K), defines the low-energy part of the infrared (IR) spectral range and partly overlaps with the far-IR (FIR) regime. For long time, this frequency range was hardly accessible by conventional techniques and, hence, was termed terahertz gap and on laboratory scale mainly accessible by elaborate backward-wave oscillator techniques.[1] During the last decades, based on the coherent and time-resolved detection of electric fields generated by ultrashort radiation bursts using photoconductive switches, this technique has taken an extraordinary development and nowadays is known as terahertz time-domain spectroscopy (THz-TDS).[2] In solid-state physics it has proven extremely useful in characterizing a variety of magnetic and vibrational low-frequency excitations, well below the canonical spectra of phonons and magnons, and has demonstrated enormous importance for characterizing weak intermolecular bonds in complex bio derived matter,[2] which is the focus of the present work. As a non-destructive radiation, it is an efficient and save way to perform full-body screening.[3] THz spectroscopy was proposed to serve in quality control and monitoring in food and pharmaceutical industries.[4] In addition, THz techniques are being developed for standoff detection of drugs or explosives,[5] while biological applications focus on tissue imaging and protein studies.[6]

Amino acids, which are linked together by peptide bonds to form polypeptide chains, are the building blocks of proteins, which are essential for life. In the zwitterionic state, α amino acids have a common structure, consisting of a central carbon atom, called α carbon, connected to a charged amino group (NH$_3^+$), a carboxyl group (COO$^-$), a hydrogen atom, and a variable side chain R. The side group determines the different chemical properties and the specific biological role of each amino acid. Amino acids can be viewed as charged molecules that crystallize from solution containing dipolar ions. 20 amino acids are found in proteins, and 22 amino acids are encoded by the genetic code. Despite their outstanding importance in the evolution of life on earth, it remains unclear if they were synthesized on primitive earth or were transported to our home planet by meteorites.[7,8,9]

The optical spectra of amino acids consist of two groups of excitations: intramolecular and intermolecular. The intramolecular excitations are mainly characteristic of the covalent bonds of the molecule constituents and can be expected in the IR or FIR regimes. The eigenfrequencies of intermolecular excitations, characteristic for crystalline materials, like translational phonon modes or collective librational excitations, are determined by significantly weaker bonds between the molecules, i.e., by van der Waals (vdW) or hydrogen bonds. In addition, the structure-forming units in amino acid crystals, i.e., the amino as well as the carboxyl groups, and the different side chains exhibit enhanced flexibility and head-to-tail motions. The corresponding excitations are highly sensitive to conformation, structure, and environment of the molecule. These excitations also often fall into the THz frequency range, and thus can ideally be mapped by THz-TDS.

Since more than 70 years, vibrational spectroscopy on crystalline amino acids is in the focus of research on bioderived matter. This is related to the fact that amino acids are building blocks of peptides and proteins, and vibrational spectra of amino acids may serve as reference spectra for more complex molecules. IR spectroscopy on amino acids started with the work of Klotz and Green[10] and Thompson *et al.*[11] Early Raman spectroscopy was initiated and documented in a series of papers by Edsall and coworkers, see, e.g., Refs. 12 and 13. THz spectroscopy on biological samples, with its ability to monitor real and imaginary part of the optical constants and its relatively high transmission through a variety of materials, combined with little damage even of fragile biological matter, started in the early 21$^{st}$ century and largely helped to explore the hardly accessible frequency range between the FIR and the microwave regime. Some early reviews can be found in Refs. 6 and 14. In the following, we will refer mainly to FIR results covering the frequency range below 120 cm$^{-1}$ and on published THz results, in both cases focusing on work including results for L-serine (Ser) and L-cysteine (Cys), which are the topical materials of this work. A comparison with Raman and inelastic neutron scattering (INS) results and with the results of model calculations will be included in the discussion section.

In the present work, we focus on the temperature evolution of the low-frequency excitation spectra of two specific amino acids: We report on Ser ($C_3H_7NO_3$) and Cys ($C_3H_7NO_2S$) both exhibiting a very similar structure, where in the side chain the thiol group ($-$ SH) in Cys is replaced by a hydroxy group ($-$ OH) in Ser. Hence, only one oxygen is substituted by sulfur, however, being responsible for a significantly different bonding behavior in the crystalline structure. The thiol group probably is the most reactive group found in amino acids. Both amino acids crystallize in the orthorhombic $P2_12_12_1$ structure with four molecules per unit cell. In crystalline form, both amino acids are in their zwitterionic form, with the proton of the carboxyl group attached to the amino group (see Figure 1), yielding a dipole moment between the negatively charged COO$^-$ and the positively charged NH$_3^+$. The molar masses of Ser and Cys are $m = 105.09$ g/mol and 121.16 g/mol, respectively. The lattice constants of the orthorhombic unit cell are $a = 8.599$, $b = 9.348$, and $c = 5.618$ Å for Ser[15] and $a = 8.144$, $b = 11.937$, and $c = 5.416$ Å for Cys.[16] Notably, while the $a$ and $c$ lattice constants are slightly smaller in Cys, $b$ is significantly larger, resulting in an enhanced volume for Cys with a significantly reduced density $\rho$ of 1.3 g/cm$^3$ compared to $\rho = 1.6$ g/cm$^3$ for Ser. The significantly less-denser crystalline packing of Cys is important for a hand-waving interpretation of the excess heat capacity and soft excitations, which will be given below and probably also for an enhanced flexibility of the side chain. On cooling from room temperature, Cys undergoes an order-disorder type phase transition,[16,17] while Ser reveals no structural anomalies below room temperature down to the lowest temperatures. To characterize the thermodynamic evolution of the two amino acids investigated, we performed temperature-dependent heat-capacity experiments from 2 K up to room temperature, where we indeed identified a clear anomaly close to 77.5 K in Cys, while the heat capacity of Ser evolves continuously from the lowest to the highest temperatures.

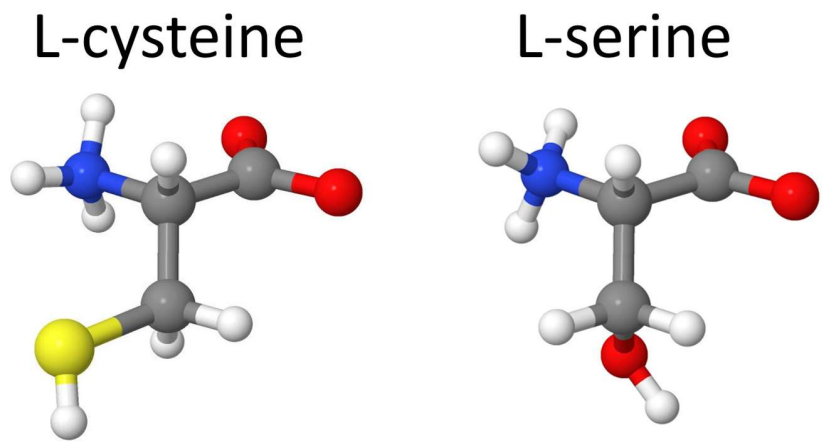

**FIG. 1.** Molecular structure of the amino acids Cys and Ser in their zwitterionic form with negatively charged $NH_3^-$ and positively charged $COO^+$. Grey spheres: carbon, white spheres: hydrogen, red spheres: oxygen, blue spheres: nitrogen, and yellow spheres: sulfur.

The main part of this work describes detailed investigations of temperature-dependent excitation spectra in both compounds for wavenumbers ranging from 20 to 115 $cm^{-1}$ and temperatures from 4 up to 300 K. The excitation spectra are consistently fitted and carefully analyzed with respect to eigenfrequencies, damping, and dipolar strength. The detailed analysis of the temperature dependence of these fundamental parameters describing the eigenmodes also will help to conclude about the microscopic nature, i.e., the eigenvectors of the observed excitations, which describe the specific displacement pattern of the atoms in the crystalline structure. Vice versa, the detailed analysis of the temperature dependence of the THz spectra in this work will allow to refine parameters of theoretical models, like bonding parameters, coupling strength, potential energy distributions, molecular force fields, or structural details of molecular conformations in the crystalline network.

We are aware that several THz and FIR investigations of excitations in amino acids, specifically covering Ser and Cys, overlapping with the frequency regime investigated in this work, exist. However, in many cases they are limited to one or two specific temperatures, mainly to room and liquid nitrogen temperature. Moreover, they lack a detailed analysis of eigenfrequencies, damping, and strength, and ignore possible temperature-induced phase transitions.[18,19,20,21,22,23,24,25,26,27,28,29,30,31] In addition, we profited enormously from the continuous improvement of THz-TDS spectroscopy during the last 20 years. It seems important to provide detailed and significant data on amino acids, essentially to model these low-lying excitations, which usually are hardly accessible in modern ab-initio molecular simulations. Two further important aspects of the low-frequency dynamics in amino acids must be mentioned: i) The slow dynamics investigated here is important for microscopic processes in protein folding and in the interaction with water molecules.[32,33] ii) The discovery of amino acids in meteorites fallen to earth suggests that the extraterrestrial medium can produce complex organic molecules and it seems important to survey and analyze the low-frequency atmospheric transmission window to possibly identify interstellar amino acids.[34,35]

## II. Experiments and Techniques

The amino acids investigated in the course of this work, Ser (purity > 99%), glycine (Gly; purity > 99%) and Cys (purity > 98%) were purchased by Sigma-Aldrich. The polycrystalline samples were pestled in a mortar to achieve grain sizes smaller than the wavelength used in the THz investigations ($\lambda \geq 50$ μm, corresponding to a wavenumber $\nu = 200$ cm$^{-1}$). This procedure is important to avoid additional scattering events of the microcrystalline objects constituting the ceramic samples. The powders without any additives where then pressed into pellets of 13 mm in diameter and a thickness of ~ 0.2 mm utilizing an external pressure of 900 MPa. The THz investigations presented in this work all were performed in transmission utilizing a TPS Spectra 3000 purchased from TeraView, which in principle allows measurements between 0.2 and 4 THz. A He-flow cryostat, equipped with 75 μm polypropylene windows, which can be operated in the sample chamber of the TPS Spectra 3000, allowed measurements from 4 K up to room temperature. The output of the spectrometer corresponds to a time-dependent electric field $E(t)$ and via measurement of a reference signal, $E(\omega) \exp[i\phi(\omega)]$ can be determined via Fourier transformation, with $E(\omega)$ being the amplitude and $\phi(\omega)$ the corresponding phase shift of the electric field. This fact documents that amplitude and phase can be determined, from which the complex optical constants, i.e., complex diffraction index $n^*$, the complex dynamic conductivity $\sigma^*$, or the complex dielectric permittivity $\varepsilon^* = \varepsilon' - i\varepsilon''$ can be calculated. In calculating the optical constants, several assumptions have to be made: The sample of well-defined thickness has two smooth plane-parallel surfaces perpendicular to the incoming light. The medium surrounding the sample has an index of refraction $n = 1$. The information on the phase is not always distinct and partly phase jumps of order $2\pi$ appear, which must be corrected. In the course of this work, we always will document frequency and temperature dependence of dielectric constant $\varepsilon'$ and dielectric loss $\varepsilon''$. The influence of the polypropylene windows of the cryostat as well as multiple scattering events in sample and environment can easily be corrected via reference measurements. For the analysis of the frequency-dependent complex optical constants, we used the RefFIT program developed by Alexey Kuzmenko.[36] In addition, for sample characterization and to get some information on possible structural phase transitions we performed standard heat-capacity experiments utilizing a Quantum Design physical property measurement system (PPMS) for temperatures 2 K < $T$ < 300 K.

## III. Experimental results and discussion

A. Heat Capacity

We measured the temperature dependence of the specific heat in Ser and Cys utilizing standard adiabatic techniques for temperatures between 2 K and 300 K, to obtain a thermodynamic characterization of the amino acids under investigation. These measurements allow identifying the occurrence of structural phase transitions below room temperature. In addition, the specific heat provides a rough measure of the number of the degrees of freedom involved in the temperature regime investigated, and allows an estimate of the fraction of phonon or vibrational modes contributing to the specific heat at ambient temperatures. We compare the results obtained in Ser and Cys with those of the simplest amino acid, glycine (Gly: $C_2H_5NO_2$), where the side chain consists of one proton only. The data presented in Fig. 2 are partly more detailed and extend to lower temperatures than previous measurements, but overall are in reasonable agreement with published results.[37,38,39,40,41,42,43] Specifically, the results are in good agreement for temperatures < 100 K, but slightly deviate at higher temperatures. Comparing them with those of the most detailed heat-capacity investigation published so far and also discussing previous heat capacity results,[42] we find deviations of the order of 5% close to room temperature, with our experiments indicating lower heat capacities. These deviations could signal differences in sample purity, different water contents, or different equilibration conditions of the ceramic samples, specifically at ambient temperatures.

A rough inspection of the raw data provides some interesting details: i) Cys reveals a clear heat-capacity anomaly close to 77.5 K, indicative of a structural phase transition, which has been reported earlier using x-ray diffraction,[16] thermodynamic,[40] and Raman-scattering experiments.[17] From heat-capacity experiments, Paukov $et\ al.$[40] reported the appearance of an extended phase transition in Cys close to 70 K, significantly below our value, while Ishikawa $et\ al.$[43] observed a sharp first-order phase transition at 76 K, close to the transition temperature reported here. The similar steep decrease of the heat capacities of Cys and Ser towards low temperatures signals similar Debye temperatures of these compounds, both significantly lower than that of Gly, corresponding to an enhanced vibrational density of states at lower frequencies in the former compounds.

Finally, the room temperature data allow a comparison with the calculated value of the high-temperature heat capacity assuming that $C\ =\ N\ 3R$, where $R$ = 8.314 J/(mol K) is the gas constant and $N$ defines the number of internal degrees of freedom contributing to the heat capacity. Cys and Ser, both contain 14 atoms/molecule, while Gly contains 10 atoms/molecule. Hence, one would expect that the high-temperature heat capacities of Cys and Ser asymptotically reach 349.1 J/(mol K), while the high-temperature value of Gly should amount 249.4 J/(mol K). Calculating the heat capacity values of the amino acids investigated, we find values of ~ 120 J/(mol K) for Cys and Ser at room temperature, documenting that only 34% of the total degrees of freedom are excited. The experimental value of Gly at room temperature corresponds to 93 J/(mol K), indicating that a similar fraction of ~ 37% of all degrees of freedom is excited at ambient temperatures. This fact is not astonishing, as it will be documented later that the excitation energies of external and internal modes in the amino acids at least span an energy region almost up to 3500 cm$^{-1}$ corresponding to temperatures of ~ 5000 K.[44] Hence, it seems clear that at room temperature only a fraction of mainly external modes will be excited.

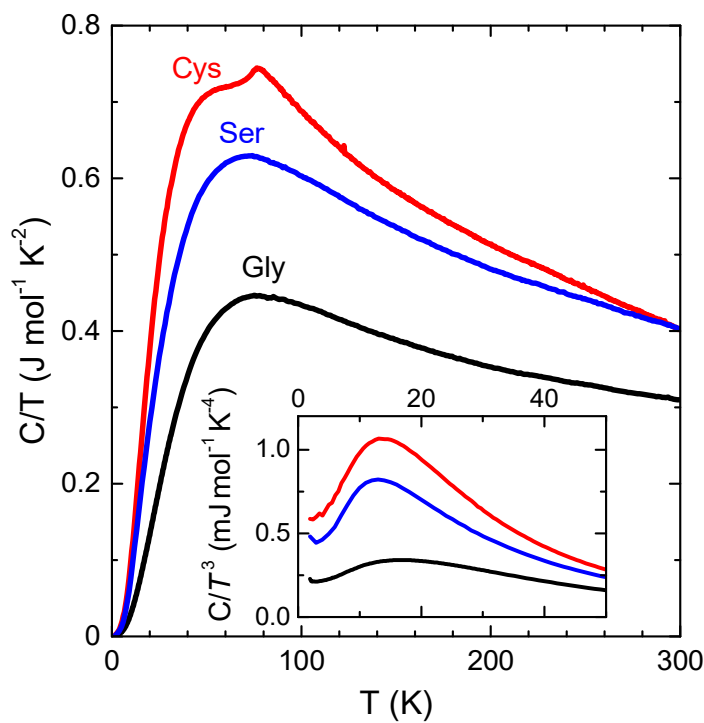

**FIG. 2.** Temperature dependence of the molar specific heat $C$ of Cys, Ser, and Gly plotted as $C/T$. The inset shows the low-temperature heat capacity, plotted as $C/T^3$ vs. $T$, to document significant excess contributions to a Debye-type behavior expected for canonical crystalline solids.

In addition, it seems interesting to have a closer look at the low-temperature regime of the specific heat. In molecular crystals one expects soft low-lying excitations corresponding to external translational or librational modes due to weak inter-molecular binding forces, like vdW forces or hydrogen bonds. These soft excitations will give rise to significant deviations from the expected low-temperature $T^3$-like temperature dependence in canonical crystals resulting from contributions of the acoustic modes only. These significant deviations from a Debye-like temperature dependence are documented in the inset of Fig. 2. A Debye behavior is characterized by a constant low-temperature plateau in $C/T^3$. At higher temperatures, values continuously decreasing on increasing temperature are expected. In the amino acids under investigation, we instead detect broad humps in the heat capacity in the 10 – 20 K range, characterizing significant excess contributions to the specific heat. Very often such behavior is considered a hallmark of glassiness and is termed boson peak.[45,46] For amino acids, this excess density of states indicates contributions from low-lying excitations, which in some cases also can be coupled to acoustic modes and by no means allows to draw conclusions concerning possible disorder in these systems. The fact that the heat capacity (Fig. 2) and excess heat capacity (inset of Fig. 2) of Cys are enhanced compared to isostructural Ser documents that the packing density of crystalline Cys is significantly lower, giving rise to lower force constants and concomitantly to an enhanced density of states at low frequencies. Ishikawa *et al*. have reported very similar observations of excess heat capacity in Cys, with experimental results in excellent agreement with our results.[43] The detection of these strong excess heat capacities in amino acids at low temperatures, as documented in the inset of Fig. 2, raises some doubts if the boson peak reported in proteins[47] results from the glassiness of the proteins investigated or just stems from low-lying vibrations of the amino acids, which are their fundamental building blocks.

The excess heat capacity is significantly larger in Ser and Cys, both with an additional sidechain, compared to Gly. This immediately signals important contributions to the low-temperature heat capacity from side-chain motions, which seem even stronger in Cys and may result from the much lower crystalline density reported above. At the lowest temperatures, from the inset of Fig. 2 one can estimate constant low-temperature values of the molar heat capacity, which allow calculating a specific-heat derived Debye temperature. The low-temperature heat capacity of a crystalline material can be derived assuming three acoustic modes only and is given by

$$C/_{T^3} = 233.8 \times R/_{\theta^3}. \tag{1}$$

With the gas constant $R$ and the Debye temperature $\Theta$. Using this naive and certainly oversimplified modelling, the Debye temperatures of Ser, Cys, and Gly can be calculated as 169, 148, and 213 K, respectively, which seem reasonable values as will be discussed later. These rather low Debye temperatures certainly correspond to the weak intermolecular binding forces in the amino acid crystals, which will be characterized by external modes, partly documented in the next section. They do not represent the strong intramolecular bindings in the amino acids, represented by high-frequency internal, i.e., bending and stretching modes. Finally, the small increase seen at the lowest temperatures in $C/T^3$, best visible for Ser (inset of Fig. 2), may correspond to tunneling processes, which reveal a much weaker temperature dependence, being assumed to be almost linear in $T$ and indicating some sort of glassiness or disorder.[48,49]

B. THz spectroscopy

As stated above, under optimal conditions the THz spectrometer used for these experiments can cover the frequency regime from 0.2 to 4 THz corresponding to wavenumbers from approximately 7 to 130 cm[-1]. For comparison with published work, we present the excitation frequencies using units of wavenumbers. To directly visualize the dynamics of a material, plots of the dielectric loss $\varepsilon''$ vs. wavenumber are the most informative representation, as the locations of the peak maxima directly correspond to the eigenfrequencies of excitations. In choosing an optimized sample thickness for these experiments, we had to make a compromise to cover the complete frequency and temperature range, and in addition, to get information on very weak and as well as on very strong excitations at the same time. This compromise resulted in the problem that close to resonance the strongest modes with the highest absorption, in some cases were not completely resolved at the lowest temperatures. However, as will be documented, a reliable estimate of the eigenfrequencies, damping, and optical weight is always possible. At this point it has to be mentioned that possibly, even at low temperatures modes with very low optical weight may be observed in thick samples with higher optical absorption only.[50]

The frequency dependence of the complex dielectric permittivity of Ser for a series of temperatures between 10 K and ambient temperature is shown in Fig. 3. Here we document the real [Fig. 3(a)] and the imaginary part of the dielectric permittivity [Fig. 3(b)] in the frequency regime accessible with the spectrometer and where excitations were observed. We did not detect any excitations below 50 cm[-1] and the scatter of the experimental data became too large for frequencies beyond 112 cm[-1] due to a strongly decreasing signal-to-noise ratio. In general, the very strong temperature dependence of the complex permittivity at resonance is a characteristic feature of the spectra shown. Figure 3 documents that a clear identification of specific modes in the THz regime range is only possible at low temperatures. Close to ambient temperatures, some modes become completely suppressed or are highly overdamped and appear as residual shoulders of neighboring peaks only. In Ser we unequivocally identify four modes between 60 and 110 cm[-1], which in Fig. 3 are labelled I to IV. An assignment of the vibrational pattern of these modes and a more detailed analysis will be discussed later. Again, we point towards the fact that close to resonance at 10 K and partly at 50 K, data points are missing around the resonance frequency, because of too high absorption.

It seems worthwhile to mention that modes I and III, on increasing temperatures reveal moderate red shifts and significant broadening, as is often observed in anharmonic crystalline lattices. The temperature dependence of modes II and IV is drastically different. As best documented in the imaginary part of the dielectric permittivity [Fig. 3(b)], mode II exhibits a moderate red shift only, but loses all its optical weight and at 290 K cannot be identified anymore, even not as a shoulder submerged under the high-frequency flank of mode I. Mode IV behaves like a well-defined eigenfrequency with low damping up to 50 K. At 90 K, the damping increases significantly and mode IV seemingly loses optical weight, well documented by the frequency dependence of the real and imaginary part of the dielectric permittivity. In addition, the background towards higher frequencies continuously increases with temperature, possibly because of red shift and increasing damping of a nearby FIR absorption, which has been reported close to 113 cm⁻¹.[18] For temperatures beyond 100 K, no characteristic signatures due to resonance absorption can be identified for this mode, neither in the real nor in the imaginary part.

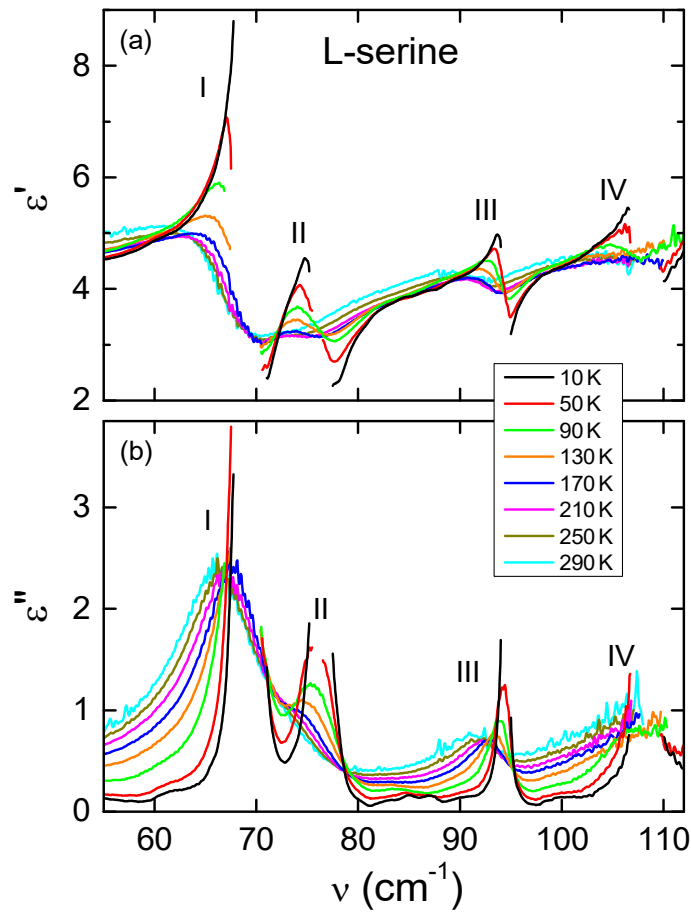

**FIG. 3.** Frequency dependence of the dielectric constant $\varepsilon'$ (a) and the dielectric loss $\varepsilon''$ (b) of Ser for temperatures between 10 and 290 K. For the strongest peaks, close to resonance data points are missing due to too high absorption for the given sample thickness. The increasing values of the complex dielectric permittivity beyond 100 cm⁻¹ signal a nearby mode with red shift and increased damping on increasing temperatures (see text).

At the lowest temperatures < 50 K, three very weak peaks can be identified in the loss spectra close to 85, 87, and 99 cm$^{-1}$ [Fig. 3(b)]. These peaks, which were also reported by other experimenters, will be discussed later, and it is unclear whether these peaks correspond to real excitations, are impurity effects, or result from non-zero background scatter. A very small additional peak seen close to 61 cm$^{-1}$ [Fig. 2(b)] has never been reported so far and will be not discussed in the following.

Due to the appearance of a structural phase transition in sulfur containing Cys at 77.5 K (see Fig. 2), the temperature dependence of the THz spectra is expected to be even more complex. The corresponding spectra of real and imaginary part of the dielectric permittivity for a series of temperatures between 10 and 290 K are shown in Figs. 4(a) and (b), respectively. At 10 K, we observe seven clear modes in the frequency range between 40 and 120 cm$^{-1}$, which are labelled from I to VII. Again, close to resonance data points are lost because of too high absorption. Because of this effect we were not able to measure the real part beyond 100 cm$^{-1}$ and below 100 K [Fig. 4(a)], while the imaginary part behaved regularly [Fig. 4(b)].

In Cys, resonance modes are observed down to significantly lower frequencies compared to those in Ser, indicating that the lower crystalline density implies softer excitations. On increasing temperatures, modes I, II, and VI reveal minor red shifts and partly significant broadening for increasing temperature as expected for anharmonic solids. Modes III and VII become almost completely suppressed for temperatures > 130 K, a behavior that could be related to the appearance of the structural phase transition. However, as mentioned above, some specific modes also lose their optical weight in Ser, without the appearance of a phase transition in the corresponding temperature regime. For mode VII, this behavior is only documented in the loss spectra [Fig. 4(b)]. At ambient conditions, modes III and VII seem to have completely lost their optical weight and even are hardly visible as shoulders at the high-frequency flanks of neighboring modes. Mode IV undergoes extreme softening and broadening. It is shifted by almost 10 wavenumbers and appears overdamped at ambient temperatures. Mode V reveals an essentially temperature independent eigenfrequency, but becomes continuously suppressed, and at ambient temperature is visible as a weak shoulder in the background close to 80 cm$^{-1}$ only [Fig. 4(b)].

Considering the detailed temperature dependence of the permittivity spectra documented in Figs. 3 and 4, with significant frequency shifts and broadening of the eigenfrequencies and, in addition, since several eigenmodes lose all their dipolar strength on increasing temperature, a complete assignment of vibrational energies and a detailed comparison with model calculations at room temperature probably would fail. Hence, a comparison of experimentally observed eigenfrequencies of amino acids with those derived from ab-initio calculations seems to be useful only at low temperatures.

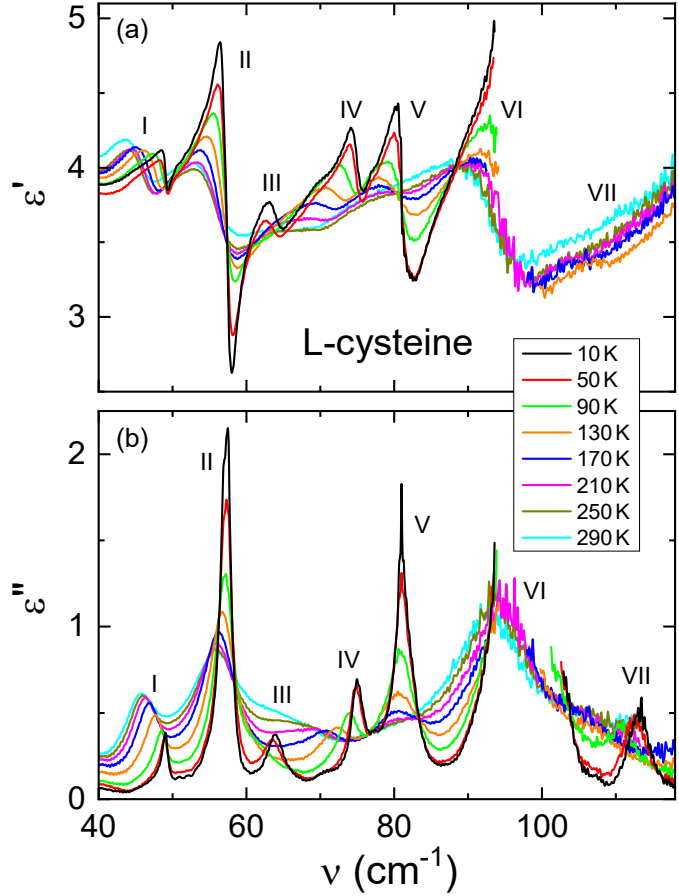

**FIG. 4.** Frequency dependence of dielectric constant $\varepsilon'$ (a) and dielectric loss $\varepsilon''$ (b) of Cys for temperatures between 10 and 290 K. For the mode labelled number VI, close to resonance absorption, data points are missing due to too high absorption for the given sample thickness. Note that, in the high-temperature phase, modes III and VII are completely suppressed, while modes IV and V are heavily overdamped.

## IV. Analysis and Discussion of THz spectra

### A. Temperature dependence of the eigenmodes

In the following, we provide a detailed analysis of the real and imaginary part of the THz results in terms of the complex dielectric permittivity, precisely determine the temperature dependence of eigenfrequencies, damping and dipolar strengths of all eigenmodes observed, and finally try to identify the translational, librational, or torsional nature of these eigenmodes. This interpretation will be additionally supported by comparison with previous experimental THz, FIR, and Raman results and with existing detailed ab-initio model calculations of eigenfrequencies of crystalline amino acids.

In crystals with space group $P2_12_12_1$, IR-active modes also are Raman active. Amino acids like Ser or Cys, with 14 atoms per molecule and 4 molecules per unit cell, should exhibit 36 internal, 3 rotational, and 3 translational modes, which should split into four branches (or exhibit a four-fold degeneracy) in the crystalline solid. Intramolecular excitations mainly will be covered by the canonical IR regime for wavenumbers $> 400$ cm$^{-1}$ and usually can be identified comparing observed spectra with tabulated eigenfrequencies, in most cases determined by measurements on isolated molecules. The THz regime mainly covers intermolecular excitations, with the bonding of molecules characterized by weak vdW forces or hydrogen bonds. In most cases, these low-frequency excitations represent coupled modes of translational and rotational motions including torsional vibrations of functional side groups. In all these cases, any identification of the displacement patterns (or strictly speaking the atomic eigenvectors) of single molecules or molecular subgroups constituting these collective modes is complicated and, in many cases, so far has not been achieved. Early on, Husain et al.[18] reported FIR spectra of amino acids covering the frequency range from 30 – 650 cm$^{-1}$ including Ser and Cys. Later on, Matei et al.[19] published a detailed study of some amino acids for wave numbers between 10 and 650 cm$^{-1}$. Further THz work on amino acids including the topical compounds studied here were reported in Refs. 20, 21, 22, 23, 24, 26, 27, 28, 29, 30, and 31, and in the following will be compared to the results derived in the present work.

We have concomitantly fitted the real and imaginary part of the complex dielectric permittivity $\varepsilon^*$ of Ser and Cys as documented in Figs. 3 and 4 by a sum of Lorentzian peaks, including the high-frequency dielectric constant $\varepsilon_\infty$:[51]

$$\varepsilon^* = \varepsilon_\infty + \sum_i \frac{\Delta_i^2}{\nu_i^2 - \nu^2 - i\gamma_i\nu} \qquad (2)$$

$\varepsilon_\infty$ describes the sum over all contributions to the real part of the dielectric permittivity arising from all high-frequency modes at $\nu \geq 120$ cm$^{-1}$. $\Delta_i$ characterizes the dipolar oscillator strength (or the optical weight), $\nu_i$ the eigenfrequency, and $\gamma_i$ the damping parameter of a given mode $i$. The strength of the optical resonance $\Delta_i$ is related to $\Delta\varepsilon$, the difference of the dielectric constant just above and below resonance absorption, via $\Delta_i^2 = \Delta\varepsilon \, \nu_i^2$.[51] For the case of phonons in ionic crystals, $\Delta_i$ is called effective ionic plasma frequency and defines the splitting of transverse and longitudinal optical phonons induced by dynamic dipoles driven by the electromagnetic wave. The damping constant corresponds to the half-width at half-maximum of the Lorentzian-shaped loss peaks. For selected temperatures, fits with Eq. (2) are indicated by the lines in Figs. 5 and 6 for Ser and Cys, respectively, documenting the suitability of the fit procedure, as explained in detail in the following.

Figures 5(a) and (b) show the obtained results for Ser for wavenumbers from 35 to 112 cm$^{-1}$ and at selected temperatures between 4 and 300 K. The mentioned weak structures close to 85, 87, and 99 cm$^{-1}$ that are visible in the low-temperature spectra of Fig. 3b, have been neglected in the shown fits. At 4 K and in the frequency regime investigated, the sum of Lorentzian fits (Eq. 2) describes real and imaginary part of Ser very well. For the eigenmodes I – IV we find eigenfrequencies of 68.9, 76.1, 94.2, and 108.5 cm$^{-1}$, respectively. However, as already documented in Fig. 3, the spectra considerably change on increasing temperatures. To achieve good fits at elevated temperatures, in addition to these four eigenfrequencies we had to assume a high-frequency mode, fixed at 113 cm$^{-1}$, which is not further discussed and was identified in early FIR measurements.[18] This mode is responsible for the increasing complex dielectric permittivity beyond frequencies of 110 cm$^{-1}$, which is best documented in the dielectric loss for 70 K. Furthermore, for temperatures above 60 K a broad and heavily smeared-out "background" Lorentzian had to be assumed to account for the increasing loss at low frequencies, which is best visible at ambient temperature for frequencies < 50 cm$^{-1}$. This additional loss peak is approximately centered around 50 cm$^{-1}$ and increases on increasing temperature. In the loss spectra, its peak maximum approximately reaches 0.5 at ambient temperatures, with a width parameter of 20 cm$^{-1}$, much broader than the characteristic eigenmodes. The origin of this feature remains unclear, but it is needed to achieve good fits, specifically to the dielectric loss at higher temperatures. For the real part of the dielectric permittivity, this additional background contribution remains insignificant. With these additional assumptions, the fits describe temperature and frequency dependence of the complex dielectric spectra of Ser very well.

As discussed earlier, modes I and III undergo moderate softening and significant broadening upon heating, but still can be identified at 300 K, while mode II exhibits a moderate red shift only and becomes increasingly suppressed. At 300 K even a shoulder-like structure on the high-frequency flank of the neighboring mode I cannot be identified anymore. Significant loss of optical weight and strong damping also appears for mode IV. Its resonance frequency can easily be identified for temperatures < 100 K. However, for a temperature window from 100 – 200 K (cf. Fig. 3) the complex dielectric permittivity cannot be fitted anymore, and both, real and imaginary part exhibit a continuous increase on increasing frequency. Interestingly, for $T$ > 200 K, the characteristic excitation-type frequency dependence reappears again and allows identifying the eigenfrequency close to 106 cm$^{-1}$. This is best documented in Fig. 5 for experimental data and fit at 300 K, where both, real and imaginary part provide evidence for dispersion effects typical for a resonance. Finally, interestingly that the dielectric constant $\varepsilon'$ of Ser measured below 40 cm$^{-1}$ is approximately 4 at the lowest temperatures and slightly increases on increasing temperature, as is usually observed in anharmonic crystals.

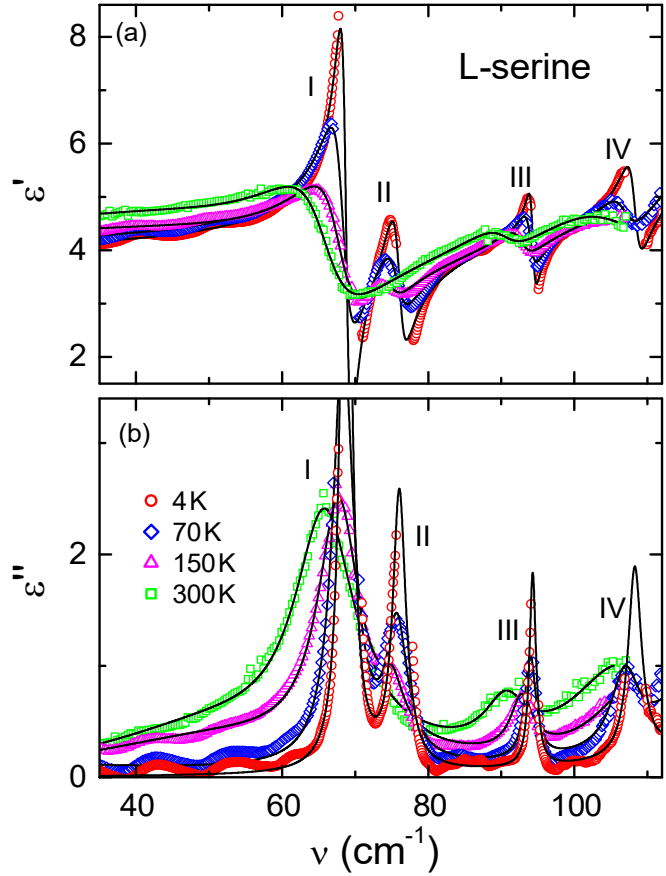

**FIG. 5.** Temperature dependence of (a) the dielectric constant $\varepsilon'$ and (b) the dielectric loss $\varepsilon''$ of Ser vs wavenumber $\nu$ at selected temperatures between 4 and 300 K. The lines represent concomitant fits of real and imaginary part of the dielectric permittivity using Lorentzian line shapes, Eq. (2) (see text). Partly, data points are missing close to resonance absorption, resulting from a too high absorption (see text).

The eigenfrequencies of the modes observed in Ser in the course of this work at 10 K and 300 K are given in Table I and are compared to published results, which in most cases were measured at higher temperatures as indicated in the table. For comparison, we also qualitatively indicated the strength of the observed peaks, as often has been done in earlier work, where the peak profiles have not been fitted and the peak height was only qualitatively assigned. Considering the different temperatures, Table I documents that there is reasonable reproducibility of eigenmodes observed in polycrystalline Ser using different experimental setups and different samples, probably with different purity and grain size. This is true for the eigenmodes I, II, and III at 10 K, all exhibiting considerable optical strength. However, mode II is completely suppressed at ambient temperature. There is some ambiguity in the frequency of mode IV, which at the lowest temperatures appears at 108.5 cm$^{-1}$, vanishes in an intermediate temperature regime, and reappears at ambient temperature close to 106 cm$^{-1}$. This mode probably corresponds to the weak shoulder reported by Husain *et al.*[18] at 106 cm$^{-1}$ and 100 K. It remains unclear whether the very weak loss peaks, identified close to 85, 87, and 99 cm$^{-1}$ in Fig. 2 at 10 K, indeed correspond to eigenfrequencies or just result from background or defect-scattering. Very weak loss peaks close to 85 cm$^{-1}$ were also reported by Korter et al.[22] and by Williams et al.[27]. In a recent THz work by Sanders et al.[31] similar weak absorption peaks located at 85.2 and 87.8 cm$^{-1}$ were identified at 6 K (see Table I). These modes only can be observed at the lowest temperatures, and they have not been considered in the fits documented in Fig. 4.

**TABLE I.** Eigenfrequencies in cm$^{-1}$ of Ser as observed by THz spectroscopy at 10 and 300 K, compared to previous THz and FIR results: Husain et al.[18], Korter et al.[22], Wang et al.[26], Williams et al.[27], Hufnagle et al.[30], and Sanders et al.[31] The temperatures where the eigenfrequencies were measured are indicated on top of the table. 77 K corresponds to LN$_2$ temperature. For comparison with earlier work, we provided a qualitative assessment of the peak strength deduced from Fig. 3: strong (s), moderate (m), weak (w), very (v), and shoulder (sh). The three vw modes close to 85, 87, and 99 cm$^{-1}$, which can be identified in the raw data at the lowest temperatures (Fig. 3b) are not included in the labelling of the peaks and were not analyzed in the present fit routine (see text).

| mode | this work (10 K) | this work (300 K) | Ref. 18 (100 K) | Ref. 22 (77 K) | Ref. 26 (RT) | Ref. 27 (77 K) | Ref. 30 (77 K) | Ref. 31 (6 K) |
|---|---|---|---|---|---|---|---|---|
| I | 68.9 (s) | 66.0 (s) | 68 (m) | 67 (s) | 67 (m) | 67 (s) | 69.4 (s) | 69.6 (s) |
| II | 76.1 (s) | | 76 (w) | 74 (w) | 80 (sh) | 73 (w) | 76.5 (m) | 78.2 (m) |
| | 85 (vw) | | | 85 (sh) | | 86 (sh) | | 85.2 (vw) |
| | 87 (vw) | | | | | | | 87.8 (vw) |
| III | 94.2 (s) | 90.7 (m) | 94 (w) | 90 (w) | 90.5 (s) | 90 (m) | 94 (m) | 95.4 (m) |
| | 99 (vw) | | | 98 (m) | | 98 (m) | | |
| IV | 108.5 (s) | 106 (w) | 106 (sh) | 102 (m) | | 103 (m) | | |

However, the very good overall reproducibility of the strongest peaks I and III, documented in Table I, with a scatter of less than 2 cm$^{-1}$ at a given temperature, is a promising result with respect to the use of THz spectra as fingerprint probes for specific amino acids. These strongest peaks even can be identified at ambient temperature, with minor redshifts. There is some ambiguity in the temperature dependence of mode II, which in our measurements is strongly suppressed at ambient conditions. In Table I we do not discuss the FIR work by Matei et al.[19] These authors reported the observation of one strong peak close to 64 cm$^{-1}$ at room temperature, in reasonable agreement with our result. In addition, they reported the observation of a weak absorption line close to 38.6 cm$^{-1}$, which is not observed in our study (see Fig. 4) and was not reported in other studies on Ser. We also did not include the results performed by Nishizawa et al.[21] at ambient conditions. Reporting on a series of amino acids at room temperature by graphs only, these authors documented two peaks close to 68 and 93 cm$^{-1}$ in Ser, in reasonable agreement with the present work. By comparing pure and racemic Ser, including a detailed comparison with model calculations, King et al.[28] focused on two lines at 67 and 79 cm$^{-1}$ at liquid nitrogen temperatures, which are close to eigenfrequencies of modes I and II reported in the present work.

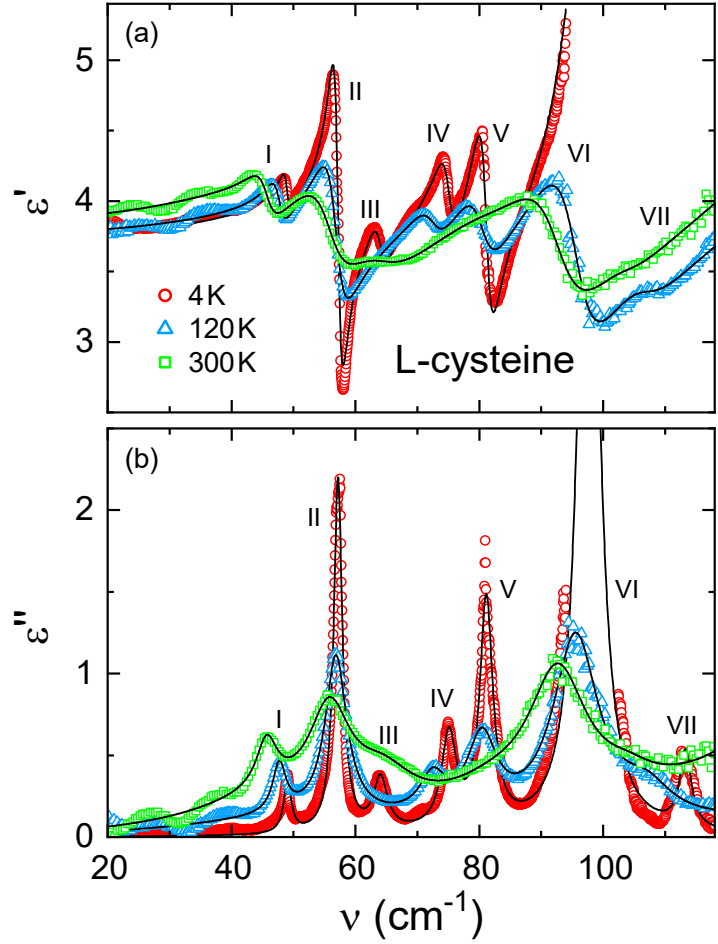

**FIG. 6.** Temperature dependence of the dielectric constant $\varepsilon'$ (a) and the dielectric loss $\varepsilon''$ (b) of Cys vs wavenumber $\nu$ for selected temperatures of 4, 120, and 300 K. The lines represent concomitant fits of real and imaginary part of the dielectric permittivity using Lorentzian line shapes, Eq. (2) (see text). Partly, data points are missing close to resonance absorption resulting from a too high absorption (see text).

The lines in Fig. 6 show fits with Eq. (2) of the frequency dependence of real (a) and imaginary part (b) of the dielectric permittivity of Cys at selected temperatures. Even though, due to too high absorption, data points are missing, from the spectra a unique determination of the mode frequencies even at the lowest temperatures is still possible (lines in Fig. 6b). At 4 K we observe the seven well-defined eigenmodes discussed above, best documented in the loss spectrum (Fig. 6b). At the lowest temperatures (< 100 K) the sequence of seven Lorentzians describes the observed spectra very well. At higher temperatures, we had to assume a further high-frequency mode to model the increase beyond 110 cm$^{-1}$ in the real as well as in the imaginary part of the dielectric permittivity. In addition, we had to assume a moderately increasing sloping background best seen in the loss spectra at ambient temperatures (Fig. 6b). Similar nearly constant-loss contributions were previously identified in a disordered crystal,[52,53] and could result from hopping conduction of charge carriers,[54,55] which may be defect electrons or protons in the present case. With these additional assumptions, the fits well describe the frequency dependence of real and imaginary part of the dielectric permittivity within experimental uncertainty. The partly wavy background, revealed especially at low frequencies, stems from weak multiple reflections at the plane-parallel surfaces of the sample, which could not be perfectly accounted for in the evaluation procedure.

As discussed above, on increasing temperatures, all modes undergo significant red shifts and damping, and, in addition, partly exhibit significant changes of their oscillator strengths. Focusing on the loss spectra (Fig. 6b) at ambient temperatures, only three strongly broadened eigenmodes persist (modes I, II, and VI), with 2 weak shoulders close to 62 and 105 cm[-1]. These shoulders are the remainders of peaks IV and VII at 4 K (see Fig. 4b). Consequently, at elevated temperatures modes III and V were not included in the fits. The good agreement with the experimental results documents that peaks III and V - within experimental uncertainty - become fully suppressed at 300 K, indicating that these eigenmodes lose all their optical weight at ambient temperatures in the high-temperature phase. Interestingly, the dielectric constant of Cys at the lowest frequencies is of order 3.8, slightly lower than that of Ser. This lower value signals a reduced over-all optical weight of the eigenmodes, and, in addition, mirrors the significantly lower mass density of Cys, with fewer charges per unit volume, both effects resulting in a lower high-frequency dielectric constant.

**TABLE 2.** Eigenfrequencies in cm[-1] of Cys as observed by THz spectroscopy at 10 and 300 K in the present investigation, compared to previous THz results: Yamamoto *et al.*[20], Nishizawa *et al.*[21], Ueno *et al.*[23], Rungsawang *et al.*[24], Brandt *et al.*[25], and Franz *et al.*[29]. Probing temperatures are indicated in the first line. For comparison with earlier work, we provided a qualitative analysis of the peak strength as deduced from Fig. 3: strong (s), moderate (m), weak (w), and shoulder (sh). The values or Ref. 24 were read off from the powder spectra at 77 K.

| mode | this work (4 K) | this work (300 K) | Ref. 20 (RT) | Ref. 21 (RT) | Ref. 23 (RT) | Ref. 24 (77 K) | Ref. 25 (RT) | Ref. 29 (20 K) |
|---|---|---|---|---|---|---|---|---|
| I | 49.0 (w) | 45.8 (w) | 45.1 (w) | 47 (w) | 46 (w) | 48 (m) | 43 (w) | 48 (m) |
| II | 57.3 (s) | 56.0 (s) | 54.7 (s) | 57 (m) | 57 (m) | 57.5 (s) | 55(s) | 58 (s) |
| III | 64.1 (w) | | | | 67 (sh) | | 66 (w) | 62 (sh) |
| IV | 73.4 (w) | 62 (sh) | | | | 73 (w) | | |
| V | 81.5 (s) | | | | | 81 (m) | | |
| VI | 97.3 (s) | 92.9 (s) | | | 93.5 (s) | 96.5 (s) | 90 (m) | |
| VII | 112.5 (m) | 105 (sh) | | | | | 102 (s) | |

The eigenfrequencies observed in Cys at 4 and 300 K in the present work are listed in Table 2 and compared to published THz work. From these results, we conclude that the strong peaks II and VI are easily reproducible even at ambient temperature, however with significant damping. This is also true for the weak peak I, which, on increasing temperature reveals minor red shift only and exhibits almost constant optical weight. Excitations I, II, and VI have partly been observed in previous investigations at room temperature,[20,21,23,25] revealing good agreement with the present results. The weak excitations I and III, which can only be observed at the lowest temperatures,[24] are in agreement with those reported by Franz *et al.*[29] at 20 K.

B. Temperature dependence of eigenfrequencies, dipolar strength, and damping

      In the following, we provide a detailed analysis of the temperature dependence of the key parameters of the observed eigenmodes to gain insight into anharmonic contributions to eigenfrequency and damping and into the temperature-dependent evolution of the optical weight. From this analysis, we try to derive some information on the nature of the translational, librational, or torsional patterns of the observed eigenmodes. In the case of Cys, we aim to document the influence of its structural phase transition at 77.5 K on the temperature dependence of the observed modes, which was neglected in most of the publications so far. For further identification of the nature of specific modes, we additionally will refer to Raman and neutron-scattering results, and to model calculations of eigenfrequencies derived from ab-initio simulations.

      Figure 7 shows the temperature dependence of eigenfrequencies (a), oscillator strength (b), and damping (c) in Ser. Here, (a) presents the normalized temperature dependence of eigenfrequencies of mode $i$, $v_i(T)/v_i(4$ K). We show normalized eigenfrequencies for better comparison of the observed red shifts and to possibly elucidate the nature of a specific eigenmode. Neglecting the rather weak anharmonic temperature dependence of the eigenmodes, in canonical ionic crystals the ionic plasma frequency and, hence, the oscillator strength (cf. discussion in section IV.A) is expected to be temperature independent. In complex molecular systems with internal and external modes, translational and rotational excitations and, in addition with hydrogen bonding and a complex network of interconnected ions, the oscillator strength may exhibit significant temperature dependence. For example, as observed by Raman scattering in Ser by Kolesov and Boldyreva[56] as well as by Bordallo *et al.*,[57] an intensity loss of specific eigenmodes on increasing temperatures was attributed to a transition from a restricted small-angle librational motion to an almost free rotation of the amino group. In this case it seems that an intensity at a given eigenfrequency is shifted towards a relaxational mode centered around zero frequency. Another possibility to explain a temperature dependence of the effective oscillator strength would be the assumption of changes of the effective charges of the fluctuating dipoles via changes in the bonding conditions of the molecules in the crystal.

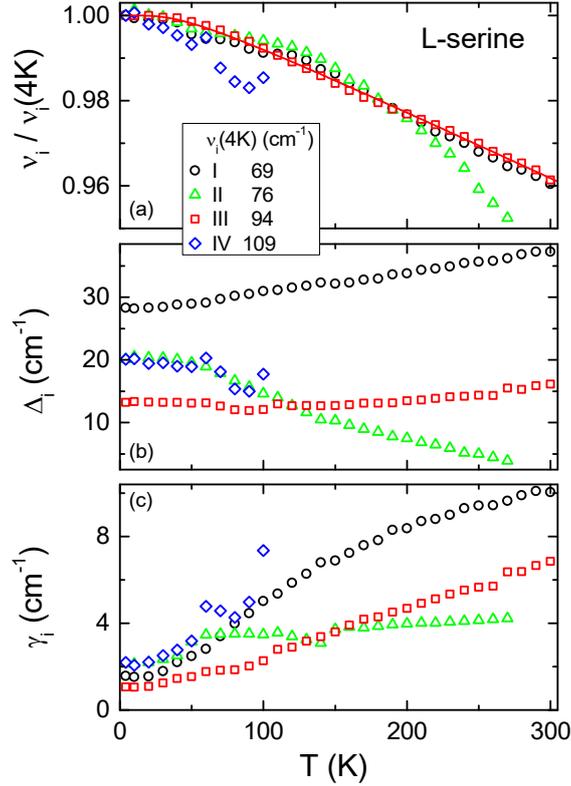

**FIG. 7.** Temperature dependence of (a) the normalized eigenfrequency $\nu_i(T)/\nu_i(4\ \text{K})$, (b) the oscillator strength $\Delta_i$, and (c) the damping constant $\gamma_i$ derived from the Lorentzian fits to the modes I – IV observed in Ser. The line in (a) is a fit of the temperature-dependent normalized eigenfrequencies of mode III (Eq. (3), see text).

Figure 7 confirms that, from the total of four modes observed in Ser at the lowest temperatures, modes I and III can be identified in the complete temperature range, while modes II and IV continuously lose optical weight [Fig. 5(b)] and are hardly observable at higher temperatures (see also Fig. 3). Modes I and III exhibit the canonical temperature dependence of eigenmodes in anharmonic crystals, and certainly can be characterized as external, i.e., intermolecular modes, involving coupled translational and librational motions of the complete molecule. As documented in Fig. 7(a), their eigenfrequencies level-off towards a constant low-temperature value and reveal a linear decrease (red shift) with increasing temperatures. The oscillator strength [Fig. 7(b)] slightly increases with temperature, however with no dramatic changes, and the damping [Fig. 6(c)] significantly increases as usually observed in anharmonic crystals. On increasing temperatures, the normalized decrease of eigenfrequencies $\nu_i$ in anharmonic crystals can be calculated utilizing the Debye temperature via[51]

$$1 - {}^{\nu_i}\!/_{\nu_{i0}} = {}^{C}\!\Big/_{\left(e^{\theta/T} - 1\right)} \qquad (3)$$

Here $\nu_{io}$ corresponds to the eigenfrequency of a specific mode at 0 K, $C$ is a constant describing the strength of the anharmonic interactions, and $\Theta$ represents the characteristic Debye temperature as defined previously in Chapter III.A. The line in Fig. 7(a) represents a characteristic anharmonic temperature dependence: Utilizing Eq. (3), mode III was fitted with a Debye temperature $\theta$ = 122 K and an anharmonicity parameter $C$ = 0.0192, which describes the experimental results very well. Fig. 7(a) implies that a similar set of parameters also can describe the temperature dependence of mode I. This Debye temperature is of similar magnitude as deduced from the low-temperature heat capacity (Fig. 2), which was determined as 169 K. It is textbook knowledge that Debye temperatures determined by different techniques often differ significantly. A high-temperature expansion of Eq. (3) gives an estimate of the linear red shift of -1.57×10⁻⁴/K for Ser, which is enhanced by a factor of 2 compared to measurements of the red shift of low-frequency modes in α-glycine by Allen *et al.*[58]

In clear contrast to the canonical anharmonic behavior of modes I and III, excitations II and IV behave significantly different: The temperature dependence of the eigenfrequencies of these modes cannot satisfactorily be described by similar fits utilizing Eq. (3). As mentioned above, these modes also reveal a radically different temperature dependence of their optical weight [Fig. 7(b)]: Specifically, mode II loses almost all optical weight close to ambient temperatures. Furthermore, the damping of this mode increases at low temperatures and then stays roughly constant for further increasing temperatures.

The temperature dependence of the characteristic parameters observed by fits to the complex dielectric permittivity of Cys (Fig. 6) are shown in Fig. 8. Again, we document the temperature dependence of normalized eigenfrequencies (a), optical weight (b), and damping (c) of the seven modes observed at the lowest temperatures. One must consider that Cys undergoes a structural phase transition at $T_s$ = 77.5 K (Fig. 2). It is reflected in the temperature dependence of the fit parameters, which evolve rather continuously at low temperatures $T < T_s$ and for $T > 130$ K, but reveal significant scatter and anomalous temperature dependence in the intermediate temperature regime between the vertical dashed lines in Fig. 8. Most significantly, mode VI loses 30% of its optical weight (Fig. 8b) and, in addition, modes III and VII become suppressed, either when crossing this temperature region (mode III) or at slightly higher temperatures (mode VII). Obviously, symmetry changes between the low and high-temperature phase partly influence the vibrational dynamics in this regime, related to the reorientational motion of the side chain, which sets in in the high-temperature phase. While there is a distinct anomaly in the heat capacity close to 77.5 K (Fig. 2), the eigenfrequencies smoothly change in a much broader temperature range pointing towards a smeared-out phase transition when focusing on the dynamics. The modes that vanish when crossing the phase transition region reveal the strongest red shifts [Fig. 8(a)] and the strongest increase of damping [Fig. 8(c)], both indicators of strong anharmonicity close to the phase transition. Overall, the anharmonic red shifts in Cys are much stronger compared to Ser, a fact that contradicts the lower Debye temperature of the former compound as deduced from the heat capacity and points towards the existence of a low-temperature structural instability. The optical strengths of all modes, except for mode VI, remain rather constant within experimental uncertainty [Fig. 8(b)], significantly different to the observation in Ser [Fig. 7(b)]. For mode IV, revealing a rather canonical red-shift behavior, the solid line in Fig. 8(a) shows a fit to its normalized temperature-dependent eigenfrequency according to Eq. (3). The resulting parameters are $\theta = 186$ K and $C = 0.129$. The larger anharmonicity parameter corresponds to a much stronger red shift compared to Ser. The obtained Debye temperature again is of comparable magnitude to the value of 148 K estimated from the low-temperature heat capacity.

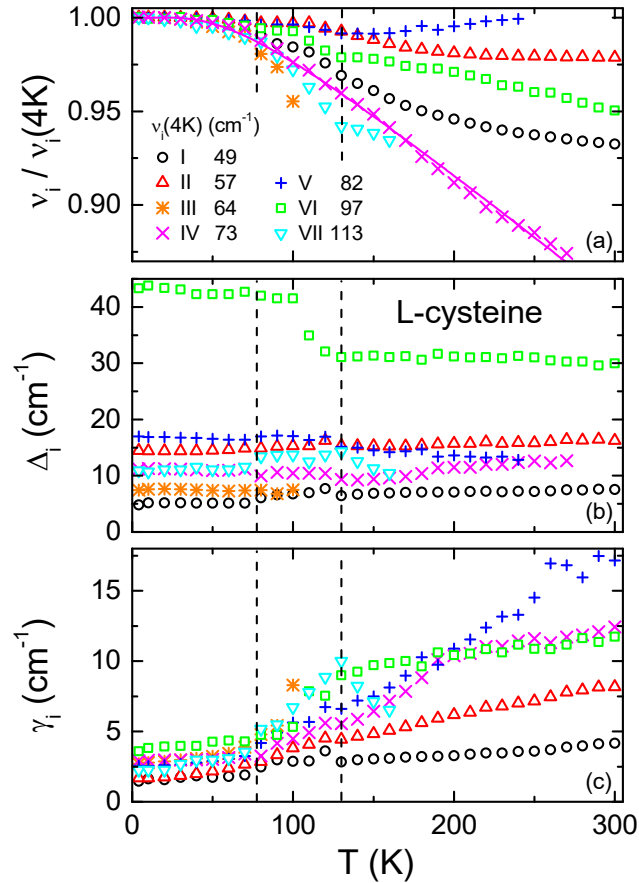

**FIG. 8.** Temperature dependence of the normalized eigenfrequency $\nu_i/\nu_i(4$ K$)$, the oscillator strength $\Delta_i$, and the damping constant $\gamma_i$ derived from the Lorentzian fits of the seven modes observed in Cys. The vertical dashed lines indicate an interval of anomalous temperature dependence of the parameters of the eigenmodes induced by the structural phase transition at 77.5 K. The solid line in (a) is a fit of the normalized eigenfrequencies of mode IV with Eq. (3).

As documented by the complex behavior in Fig. 8, an identification of the microscopic nature of the observed modes is not straightforward. Whether these modes are mainly of translational, librational, or torsional character cannot easily be deduced from the observed temperature dependences of the Lorentzian parameters. It seems even unclear to what extent these vibrations are of external or internal character. The normalized eigenfrequencies [Fig. 8(a)] of modes I, II, and VI reveal a clear change of slope, which becomes significantly reduced in the high-temperature phase. The temperature dependent eigenfrequency of mode V passes through a minimum and even increases in the high-temperature phase. Interestingly, this temperature dependence roughly mirrors the temperature dependence of the lattice constants $a$ and $b$, which exhibits minima in the 150 to 250 K range.[17] Furthermore, modes III and VII vanish at high temperatures. Only mode IV follows a canonical anharmonic decrease [line in Fig. 8(a)], however with a strongly enhanced anharmonic coefficient $C$ compared to Ser. Before comparing the observed eigenfrequencies with existing model calculations, we want to discuss the nature of the structural phase transition in Cys and to detect its influence on the temperature dependence of the measured eigenfrequencies.

The microscopic nature of the phase transition in Cys has been identified by detailed x-ray diffraction studies at 30 K by Moggach *et al.*[16] Although the crystal structure of Cys is dominated by N−H····O hydrogen bonding, the thiol group is also capable of forming hydrogen bonds. At ambient conditions, the H atom of the thiol group is disordered in such a way as to form intermolecular S−H····S and S−H····O bonds. In the low-temperature phase, the structure becomes fully ordered with the formation of S−H····S bonds only. Hence, the structural phase transition observed at 77.5 K corresponds to a canonical order-disorder phase transition involving mainly the torsional-reorientational motion of the thiol group.

## C. Comparison with IR, Raman, and neutron-scattering results, and with model calculations

Complex molecular crystals exhibit three general types of motions: translational and librational modes, which are always external, i.e., intermolecular, as well as vibrational excitations, which are predominantly internal, i.e., intramolecular. The latter are largely dominated by covalent bonds and, hence, will dominate in the FIR and IR regime; the former are governed by much weaker vdW and hydrogen bonds and, hence, will appear in the low-frequency regime. One can expect that internal and external modes are separated by a distinct frequency gap, however, these modes will not necessarily be always decoupled. The external phonon modes can be essentially characterized as rigid molecular translations and small-angle rotations. In crystalline amino acids with 4 molecules per unit cell, there will be 3 acoustic modes and 21 external optical excitations consisting of 9 translational optical modes (rigid mass optical phonons) and 12 librational excitations describing small-angle intermolecular rotational excitations around one of the three crystallographic axes. In principle, these 21 optical excitations should dominate the low-frequency regime investigated in this work. However, one must be aware that some soft torsional excitations of molecular subgroups may additionally appear in this frequency window. This number of 21 theoretically expected excitations has to be compared with the observed 4 excitations in Ser and 7 excitations in Cys. This fact documents that some of the excitations are very weak,[50] are not IR active, or lie outside the reported frequency regime.

Since the very beginning, authors tried to assign certain observed eigenfrequencies to a specific molecular motion or to specific eigenvectors of atomic displacements. However, as will be documented below, in many cases this is not an easy task. For example, Husain *et al.*[18] overall characterized all modes with eigenfrequencies below 145 cm$^{-1}$ as lattice vibrations without any further discussion. Matei *et al.*[19] proposed that all low-frequency modes are rather collective and strongly influenced by hydrogen bonds. Parker,[59] attempting to assign the vibrational spectrum of Cys, stated that all modes below 200 cm$^{-1}$ are strongly mixed and, hence, these authors disregarded the possibility to distinguish between translational, librational, or so-called low-energy "skeletal" modes[60] of the carbon backbone of the amino acids. Sanders *et al.*[61] characterized all modes observed in alanine below 100 cm$^{-1}$ as purely translational or rotational external modes.

Before discussing translational and librational motion in more detail, we have a closer look on vibrational excitations. Vibrations can be subdivided into stretching ($\nu$), bending ($\delta$: subdivided into wagging, rocking, scissoring, and twisting) or torsional excitations ($\tau$). While in most cases stretching and bending excitations can be characterized as high-frequency excitations, torsional rotations around single bonds can act as soft degrees of freedom that may strongly influence the low-frequency dynamics. There have been some attempts documented in literature to determine the energy barriers against torsional excitations of molecular groups.[62] From these estimates it seems that hindering barriers are lowest for torsional excitations of the methyl group being of order 260 meV,[62] and excitations in this rather soft potential could influence the low-frequency molecular dynamics in amino acids. We also should take into consideration that at elevated temperatures a torsional motion may transform into a reorientational motion, and finally into an almost free rotation, losing all its vibrational intensity at finite frequencies. Furthermore, the different excitation patterns of single molecular groups are strongly coupled and single molecular excitation patterns, as outlined above, will hardly exist. Hence, when assigning low-frequency modes, one aims to denote the dominant excitation of a specific mode neglecting further contributions from a variety of different displacement patterns.

In this respect, inelastic neutron-scattering experiments performed to determine the low-frequency (< 100 cm$^{-1}$) phonon dispersion in single-crystalline alanine[62] provide illuminating insight into the microscopic nature of low-lying vibrational modes in amino acids. The authors of Ref. 62 report on the observation of three acoustic and up to four optic modes along all three main symmetry directions for frequencies below 100 cm$^{-1}$. The optic modes were identified mainly as collective librational motions of the rigid molecules, however, with significant admixture of translational motions. The observation of four optical modes by INS has to be compared with the observation of 5 modes by THz spectroscopy in the similar frequency regime.[63] Crowell and Chronister[64] described the low-frequency optical modes of alanine as external and rigid molecular translations and rotations and even though the correct eigenvectors are not known, it was assumed that librational motions correspond to small-angle hindered rotations around one of the three principle crystallographic axes.

The detailed experimental information on temperature dependencies of eigenfrequencies, dipolar strength, and damping, as documented in Figs. 7 and 8, will now be used to assign the observed eigenmodes in Ser and in Cys to specific excitation patterns, i.e., to specific eigenvectors of the atomic displacements. This will be done by additionally comparing our results to published THz, FIR, Raman, and inelastic neutron-scattering results, and by specifically referring to detailed ab-initio calculations based on Hartree-Fock (HF) methods or density functional theory (DFT). In amino acids there are several ab-initio calculations on the HF or DFT level. The possible shortcomings of the calculation of low-frequency modes by ab-initio methods have recently been treated by Allen *at al*.[64] Tarakeshwar and Manogaran[65] and Tellez *et al*.[66] calculated vibrational frequencies of Ser in its non-ionized form in the single molecule. Both groups found torsional excitations in the low frequency regime $< 120$ cm$^{-1}$, but assigning these modes to different molecular groups. However, in the crystalline form, Ser is in a zwitterionic form, and a direct comparison of low-energy vibrational excitations of the neutral, non-charged single molecule may not be meaningful. Vibrational excitations of the single molecule of zwitterionic Ser were published by Chakraborty and Manogaran[67] and by Pawlukojc *et al*.[68] to support their experimental eigenfrequencies collected by Raman and neutron-scattering experiments, respectively. Korter *et al*.[22] moved one step further to support his THz results and calculated eigenfrequencies using the crystallographic coordinates and a supercell with four molecules per unit cell. Finally, and from our perspective most convincing, Williams and Heilweil[27] as well as King *et al*.[28] provided ab-initio calculations with a clear correlation of eigenfrequencies to distinct excitations patterns. The former provided an analysis of the vibrational spectra of Ser utilizing molecular dynamics (MD) simulations of Born-Oppenheimer and Car-Parrinello type assuming two unit cells with 8 molecules. They compared their MD simulations with published THz,[22] Raman,[56] and neutron-scattering results.[68] King *et al*.[28] performed DFT calculations with periodic boundary conditions.

Most of these calculated eigenmodes are characterized by complex displacement patterns with contributions from different translational and rotational excitations. By comparison with these detailed ab-initio calculations, mode I of Ser, located at 68.9 cm$^{-1}$, is described by William and Heilweil.[27] as external mode of purely translational character. King *et al*.[28] characterize this mode with 50% translational, 25% external librational character and the remaining fraction of 25% as an internal torsion of the $NH_3^+$ group. Mode III at 94.2 cm$^{-1}$ also mainly corresponds to a translational motion coupled to torsional excitations.[28] Mode II at 76.1 cm$^{-1}$, which vanishes close to ambient temperatures, corresponds to a torsional motion of the amino group coupled to translational excitations.[28] Finally, mode IV at 108.5 cm$^{-1}$ probably can be characterized as torsional motion of the $CO_2^-$ group coupled to a smaller fraction of external rotation.[27] This characterization of the nature of vibrations in Ser is in accord with the temperature dependence of the parameters shown in Fig. 7. Modes with predominant intermolecular translational or librational character behave like canonical anharmonic excitations with a continuous and moderate decrease of eigenfrequency, an increase of damping, and constant optical weight. The modes II and IV with predominant intramolecular torsional motion behave systematically different, and specifically lose optical weight on increasing temperature.

Several ab-initio MD calculations on a HF or DFT level have in the past also been performed for Cys,[20,24,25,67,68,69,70] mainly to explain some of the vibrational excitations observed in Raman, THz, and INS experiments. These ab-initio calculations document that the low-lying modes (< 120 cm$^{-1}$) studied by THz spectroscopy predominantly result from external translational and/or librational motions, involving phonon-like molecular vibrations throughout the crystal structure. However, some of the modes correspond to torsional excitations of the specific end groups, which participate in the three-dimensional network of hydrogen bonds. A closer look into the predictions of various ab-initio calculations of eigenfrequencies, performed with significantly different complexity, always involve similar classes of excitations, but do not converge to a unified view of the displacement pattern for observed specific eigenfrequencies. Hence, we again try to combine the temperature-dependent anharmonic effects observed in the course of this work with results of ab-initio calculations. Due to the increased number of excitations in the THz regime and due to the structural phase transition, involving orientational ordering of the sidechain, the analysis of the low-lying excitation spectrum of Cys is much more complex than for Ser and by no means straightforward.

Guided by the temperature dependence of eigenfrequency, damping, and optical weight, as documented in Fig. 8, it seems reasonable to assume that modes I, II, V and VI at 49, 57.3, 81.5 and 97.3 cm$^{-1}$ correspond to coupled intermolecular translational and librational excitation of the rigid Cys molecule in the crystal structure. The weak vdW or hydrogen bonding, together with the relatively large molecules, explains the low excitation frequencies. In Ser, the corresponding translational/librational modes were identified at 69 and 94.2 cm$^{-1}$. The low eigenfrequency of mode I as observed in Cys, may indicate slightly weaker intermolecular bonding corresponding to a significantly reduced mass density compared to Ser. We would like to recall that in isostructural L-alanine up to 4 optical modes have been observed at the zone center of the Brillouin zone by inelastic neutron scattering in a frequency regime from 40 - 90 cm$^{-1}$ in reasonable agreement with the present results.[62] In our interpretation, these modes correspond to the equivalent of optical phonon modes at the zone center in prototypical molecular systems with coupled translational and librational character. These modes reveal constant, temperature-independent optical weight, and a moderate increase of anharmonic damping. As discussed above, the temperature dependence of the eigenfrequency of mode V reveals a minimum and mimics the temperature dependence of the crystallographic $a$ and $b$ lattice constants,[17] probably because of the structural phase transition. This fact further supports the identification as external excitation. The translational/librational excitation at 97.3 cm$^{-1}$ (mode VI) also is influenced by the structural phase transition: it reveals rather normal red shift on heating, with a slight anomaly in the phase-transition regime, a step-like increase in damping and a step-like decrease in optical weight, both probably induced by the disordering of S − H⋯O bonding when crossing the phase-transition temperature.

There is some ambiguity in assigning the microscopic excitation pattern of the remaining modes III, IV, and VII in Cys included in Fig. 8, which probably correspond to torsional excitations of some of end or head groups or of the whole side chain. Mode III at 64.1 cm$^{-1}$, which is lost in the temperature regime of the order-disorder phase transition, probably corresponds to a torsional excitation of the side chain, which changes its nature into a reorientational motion by the disordering phenomena of the sulfur-hydrogen bonds. Mode IV at 73 cm$^{-1}$ reveals a strong red shift, a strong increase in damping, and at high temperatures is hidden under the high-frequency flank of mode II. The analogous mode II of Ser, close to 76 cm$^{-1}$, reveals a similar red shift, but in addition loses optical weight and finally also is hidden beneath the neighboring mode. We assign these modes in both compounds to the torsional excitations of the amino group. Finally, according to Chakraborty and Manogaran[67], mode VII close to 112.5 cm$^{-1}$ at low temperatures may correspond to a complex torsional excitation of the carbon backbone of Cys. In our experiments this mode loses most of its optical weight above the structural phase-transition temperature.

## V. Concluding remarks

The present work deals with low-frequency (< 120 cm$^{-1}$) excitations in the amino acids Ser and Cys as determined by THz spectroscopy. Low-frequency translational, librational, and torsional excitations in the amino acids are of specific importance and interest as they significantly contribute to the vibrational entropy, strongly influence a variety of biological processes, and may be used as fingerprints for the detection of different amino acids. Both systems belong to the group of proteinogenic α-amino acids and have a very similar structure with the side chain -CH$_2$OH in Ser being replaced by -CH$_2$SH in Cys. Both crystallize in space group $P2_12_12_1$ at ambient temperatures. For Cys, our heat-capacity experiments reveal a clear anomaly at the order-disorder structural phase transition of Cys at 77.5 K, involving the ordering of the thiol group. In low-temperature heat capacity measurements, we identify significant excess heat contributions beyond the expected Debye behavior, which are strongest in Cys and weakest in Gly, providing some insight into the low-frequency vibrational density of the compounds investigated. This low-frequency flexibility is of utmost importance for all bioderived structure-building activities at ambient temperatures.

We provide the detailed temperature dependence of eigenfrequency, damping, and optical weight of all detected low-lying modes in both amino acids from 4 K up to room temperature. Despite the structural similarity of both compounds, the vibrational patterns look significantly different, with four clearly identified eigenmodes in Ser compared to seven modes in Cys as observed at low temperatures. In both systems, the excitation spectra change considerably upon increasing temperatures. Due to the suppression of some modes at high temperatures, a comparison of the observed eigenfrequencies with those derived from ab-initio calculations is useful at low temperatures only. Finally, we analyzed the temperature evolution of eigenfrequencies, damping, and optical weight, mainly to get information on the eigenvectors of the atomic displacements describing the microscopic nature of the vibrational patterns observed. Even though it does not seem straightforward to assign the low-frequency modes to purely translational, librational, or torsional modes, based on the detailed temperature dependence and by comparison with existing ab-initio calculations, we tried to characterize the observed modes by their most dominant fraction of excitation patterns. We conclude that most of the observed eigenfrequencies in both compounds correspond to coupled external translational and librational modes of the rigid molecules in the crystallographic structure, as well as to some internal torsional vibrations of molecular end groups. These detailed measurements and analysis should allow the fine-tuning of model parameters in ab-initio calculations, which are specifically uncertain in the low-frequency regime dominated by intermolecular collective modes. In addition, due to the importance of amino acids as building blocks of life in biosynthesis and in the evolution of life on earth, it seems important to develop vibrational fingerprint measures of the different amino acids for which THz spectroscopy seems an ideal method.

## AUTHOR DECLARATIONS

Conflict of Interest
The authors have no conflicts to disclose.

## DATA AVAILABILITY

The data that support the findings of this study are available from the corresponding author upon reasonable request.